\documentclass[amsthm]{elsart}
 \pdfoutput=1

\usepackage{indentfirst}
\usepackage{url}

\usepackage{yjsco}
\usepackage{natbib}
\usepackage{amsmath}
\usepackage{amsfonts}
\usepackage{amsthm}
\usepackage{amssymb}
\usepackage{bm}
\usepackage{algpseudocode}
\algrenewcommand\algorithmicrequire{\textbf{Input: }}
\algrenewcommand\algorithmicensure{\textbf{Output: }}
\def \Res  {{\rm Res}}
\def \MCproj  {{\rm MCproj}}

\def \discrim  {{\rm discrim}}
\def \sqrfree  {{\rm sqrfree}}

\def \lc  {{\rm lc}}
\def \Bproj  {{\rm Bproj}}
\def  \zero {{\rm Zero}}
\def  \Nproj {{\tt Nproj}}

\def  \NKproj {{\tt NKproj}}
\def \Proineq {{\tt Proineq}}
\def \Bprojection {{\tt Bprojection}}
\def \Findk {{\tt Findk}}
\def \SRes {{\tt SRes}}
\def \Findinf {{\tt Findinf}}

\def \R {{\mathbb R}}
\def \Z {{\mathbb Z}}
\def \rr {{\mathcal R}}
\newcommand{\xx}{\bm{x}}
\newcommand{\A}{\bm{a}}
\newcommand{\B}{\bm{b}}

\def \FI {{\tt FI}}
\def \QEPCAD {{\tt QEPCAD}}
\def \PCAD {{\tt PCAD}}
\begin{document}

\newtheorem{ex}{\qquad Example}[section]

\begin{frontmatter}

\title{\textbf{Proving Inequalities and Solving Global Optimization Problems via Simplified CAD Projection}}
\author{Jingjun Han\corauthref{cor},}
\corauth[cor]{Corresponding author}
\ead{hanjingjunfdfz@gmail.com}
\author{Zhi Jin,} 
\ead{j2i5nzhi@sina.com}
\author{Bican Xia}
\ead{xbc@math.pku.edu.cn}
\address{LMAM $\&$ School of Mathematical Sciences, Peking University, Beijing 100871, China}
\thanks{The first two authors were supported by the President's Fund for Undergraduate Students of Peking University. The work was supported by NSFC-11290141, NSFC-11271034 and the SKLCS project SYSKF1207.}

\begin{abstract}
Let $\xx_n=(x_1,\ldots,x_n)$ and $f\in \R[\xx_n,k]$. The problem of finding all $k_0$ such that $f(\xx_n,k_0)\ge 0$ on $\mathbb{R}^n$ is considered in this paper, which obviously takes as a special case the problem of computing the global infimum
or proving the semi-definiteness of a polynomial. 
For solving the problems, we propose a simplified Brown's CAD projection operator, \Nproj, of which the projection scale is always no larger than that of Brown's. For many problems, 
the scale is much smaller than that of Brown's. As a result, the lifting phase is also simplified. Some new algorithms based on \Nproj\
for solving those problems are designed and proved to be correct. Comparison to some existing tools
on some examples is reported to illustrate the effectiveness
of our new algorithms.
\end{abstract}

\end{frontmatter}
\section{Introduction}

Let $\R$ be the field of real numbers and $\xx_n=(x_1,\ldots,x_n)$ be $n$ ordered variables. Consider the following three well-known problems.

\textbf{Problem 1.} For $f\in \R[\xx_n]$, prove or disprove $f(\xx_n)\ge0$ on $\R^n$.

\textbf{Problem 2.} For $f\in \R[\xx_n]$, find the global infimum $\inf f(\mathbb{R}^n)$.

\textbf{Problem 3.} For $f\in \R[\xx_n,k]$, find all $k_0\in \mathbb{R}$ such that $f(\xx_n,k_0)\ge 0$ on $\R^n.$


A lot of work has been done for Problem 1 since \citet{Hilbert}. For related classical
results, see for example, \citet{Bernstein,Artin,Polya,Hardy,Motzkin(1952),Motzkin(1969),Berg}. In recent years, many other methods have been proposed. See for
example, \citet{Putinar,Schweighofer,yang1,Scheiderer,Yao2,Castle,xu}.

Problem 2 can be regarded as a generalization of
Problem 1. Various methods based on different principles have been proposed for solving Problem 2,
including methods based on Gr\"{o}bner base \citep{Lindberg, Hanzon}, semi-definite programming or SOS based methods \citep{parrilo,Lasserre,Jibetean,nie,pham,guo}, and methods based on Wu's method \citep{xiao}. Semi-definite programming returns numerical solutions, which, in some cases, may be larger than the supremum.
Some methods need additional assumptions, for example, that the polynomial can attain the infimum \citep{nie} or the zero of the set of the first partial derivatives is zero-dimensional \citep{Lindberg}. \citet{Safey} provided a certified algorithm based on the topology property of generalized critical values \citep{kurdyka2000semialgebraic} to solve problem 2. 
The algorithm was designed to compute critical values and asymptotic critical values based on Gr\"{o}bner basis computation. The algorithm has been implemented in the RAGlib package of Maple.

Problem 3 is more general. It is a typical problem of quantifier elimination (QE)
on real closed fields. Algorithms of single exponential complexity to solve Problem 3 in the case of integer coefficients were given in \citep{grigor1988solving,renegar1992computational,heintz1993theoretical,Basu1996,Basu2006}. They are all based on computation of critical values and have not lead to efficient implementations.
Theoretically, it is feasible to apply general quantifier elimination methods 
\citep{collins1,collins2,collins3,Dolzmann1999} to solve Problem 3. Since the problem of QE is inherently doubly exponential in the number of variables \citep{Fischer1974,Davenport1988,brown2007}, general tools for QE are not the best choice in practice for special problems.

The original algorithm of Cylindrical Algebraic Decomposition (CAD) \citep{collins1} for QE is not efficient since the algorithm process of CAD projection phase involves a large amount of resultant calculation and the lifting phase needs to
choose a sample point in every cell. Hence a lot of work tries to improve the CAD projection.
A well known improvement is Hong's projection operator which is applicable in all cases \citep{Hong}. For many problems, a smaller projection operator given by \citet{McCallum1,McCallum2}, with an improvement by \citet{brown}, is more efficient. \citet{Strzebonski} proposed an algorithm called Generic Cylindrical Algebraic Decomposition(GCAD) for solving systems of strict polynomial inequalities, which made use of the so-called generic projection, the same projection operator as that proposed by \citet{brown}. Based on Wu's principle of finite kernel \citep{wu1998,wu2003}, Yang proposed without proof the successive resultant method \citep{yang2001,xia} to solve the global optimization problem involving polynomials and square-roots, in which Brown's projection is used in the projection phase and only sample points from the highest
dimensional cells need to be chosen in the lifting phase. \citet{McCallum3} once pointed out that, in order to prove a polynomial inequality, only those sample points from the highest dimensional cells need to be chosen. 
\citet{xiaorong} proved that, in terms of the Brown projection, at least
one sample point can be taken from every highest dimensional cell via the
Open CAD lifting. 

In this paper, we consider how to improve the CAD based methods for solving Problems 1, 2 and 3. We propose a simplified Brown's CAD projection operator, \Nproj, of which the projection scale is always no larger than that of Brown's. Some new algorithms based on \Nproj\
for solving those problems are designed and proved to be correct. Some examples that could not be solved by existing CAD based tools have been solved by our tool.

The structure of this paper is as follows. Section \ref{se:2} shows by a simple example our main idea of designing new projection operators.
Section \ref{se:3} introduces basic definitions, lemmas and concepts of CAD and Brown's projection. Section \ref{se:4} proves
the correctness of the successive resultant method proposed in \citet{yang2001}. In Section \ref{se:5} and Section \ref{se:6}, our new projection operator \Nproj\ is introduced and some new complete algorithms
based on \Nproj\ are proposed for solving the above three problems.
The correctness of our algorithms are proved.
The last section includes several examples which demonstrate the process and
effectiveness of our algorithms.

\section{Main idea}\label{se:2}

First, let us show the comparison of our new operator $\Nproj$ and Brown's projection operator on the following simple example. Formal description and proofs of our algorithms are given in subsequent sections.

\begin{ex}
Prove or disprove
\[\forall (x,y,z)\in\R^3 (f(x,y,z)\ge 0)\]
where
\[f(x,y,z)=4z^4-4z^2y^2-4z^2+4y^2x^4+4x^2y^4+8x^2y^2+5y^4+6y^2+4x^4+4x^2+1.\]

We solve this example by a CAD based method. First we apply Brown's operator and take the following steps:

Step 1. \[f_1:=\Res(\sqrfree(f),\frac{\partial}{\partial z}\sqrfree(f),z) = 1048576g_1^3g_2h_1^2h_2^2,\] where
\[g_1=y^2+1,\ g_2=4x^4+4x^2y^2+4x^2+5y^2+1,\ h_1=x^2+1,\ h_2=x^2+y^2,\]
``\Res" means the Sylvester resultant and ``\sqrfree" means ``squarefree" that is defined in Definition \ref{de:sqrfree}.

Step 2.
\[\begin{array}{rl}
 f_2 :=  & \Res(\sqrfree(f_1),\frac{\partial}{\partial y}\sqrfree(f_1),y)\\
    =  & \Res(g_1g_2h_1h_2,\frac{\partial (g_1g_2h_1h_2)}{\partial y},y)\\
    =  & 16384(x^2+1)^{15}(x-1)^{12}(x+1)^{12}(2x^2+1)^2(4x^2+5)^2x^2.
 \end{array}\]
Actually, computing $f_2$ is equivalent to computing the following $6$ resultants.
\begin{enumerate}
\item[(a)] $\Res(g_i,\frac{\partial}{\partial y}g_i,y)\ \ (i=1,2),$
\item[(b)] $\Res(h_2,\frac{\partial}{\partial y}h_2,y),$
\item[(c)] $\Res(g_1,g_2,y), \Res(g_i,h_2,y)\ \ (i=1,2).$
\end{enumerate}

Step 3. By real root isolation of $f_2=0$, choose $4$ sample points of $x$: $x_1 = -2$, $x_2 = -\frac{1}{2}$, $x_3 = \frac{1}{2}$, $x_4 = 2$.
At the lifting phase, we first get $4$ sample points of $(x,y)$ for $f_1(x_i, y)\ne 0$: $(-2,0), (-\frac{1}{2}, 0)$, $(\frac{1}{2}, 0)$, $(2, 0)$. Then get $4$ sample points of $(x,y,z)$ for $f(x_i, y_i, z)\ne 0$: $(-2,0,0), (-\frac{1}{2}, 0,0)$, $(\frac{1}{2}, 0,0)$, $(2, 0,0)$.

Step 4. Finally we should check that whether or not $f(x,y,z)\geq0$ at all the $4$ sample points. Because $f(x,y,z)\ge 0$ at all the sample points, the answer is
\[\forall (x,y,z)\in\R^3 (f(x,y,z)\ge 0).\]

Now, we apply our new projection operator to the problem. Step 1 is the same as above. According to our algorithm, at Step 2, we need only to compute the following $3$ resultants. 
\begin{enumerate}
\item[(a)] $\Res(g_i,\frac{\partial}{\partial y}g_i,y),\ \ (i=1,2)$
\item[(b)] $\Res(h_2,\frac{\partial}{\partial y}h_2,y),$
\end{enumerate}
That gives a polynomial (after squarefree) $f_2'=x(x^2+1)(2x^2+1)(4x^2+5).$

At Step 3, by real root isolation of $f_2'=0$, we choose $x_1=-1$ and $x_2=1$ as sample points for $x$. At lifting phase, compute $2$ sample points $(-1,0),(1,0)$ of $(x,y)$ and verify that $g_1g_2\ge 0$ at the two points. Then compute $2$ sample points $(-1,0,0),(1,0,0)$ of $(x,y,z)$.

At Step 4, check whether or not $f(x,y,z)\ge 0$ at all the 2 sample points. Because $f(x,y,z)\ge 0$ at all the sample points, the answer is
\[\forall (x,y,z)\in\R^3 (f(x,y,z)\ge 0).\]
\end{ex}

For this example, our new projection operator \Nproj\ avoids computing 3 resultants compared to Brown's operator. 
In general, for a polynomial $f(x_1,\ldots,x_n)\in \Z[x_1,\ldots,x_n]$, \Nproj\ first computes $f_1=\Res(\sqrfree(f),\frac{\partial}{\partial x_n}\sqrfree(f),x_n)$ as other CAD based methods do. Then, divides the irreducible factors of $f_1$ into two groups: $L_1$ and $L_2$, where $L_1$ contains all factors with odd multiplicities and $L_2$ contains all factors with even multiplicities. Compared to Brown's projection, at the next level of projection, neither the resultants of those polynomial pairs of which one is from $L_1$ and the other from $L_2$ nor the resultants of the polynomial pairs in $L_1$ are to be computed. Therefore, the scale of \Nproj\ is no larger than that of Brown's. For a wide class of problems (see for example Remark \ref{re:nproj}), especially when $n\ge3$, the scale of \Nproj\ is much smaller than that of Brown's. Based on the new operator, we obtain a new algorithm \Proineq\ (see Section \ref{se:5} for details) to prove or disprove a polynomial to be positive semi-definite. 

The main idea behind our method is that Lemma \ref{lem:weiss} provides a condition from which it can be derived that, (roughly speaking) to show that a polynomial $f(x_1,...,x_n)$ is positive semi-definite (as a polynomial in $x_n$ whose coefficients are given parametrically as polynomials in $x_1,...,x_{n-1}$) throughout a region $U$ in $(n-1)$-space it suffices that the even multiplicity factors are sign-invariant in $U$ (typical CAD) and the odd factors are semi-definite in $U$ (a weaker condition than sign-invariance). Please see Theorems \ref{thm:3} and \ref{thm:4} in Section \ref{se:5} for details.

\section{Preliminaries}\label{se:3}
In this paper, if not specified, for a positive integer $n$, $\A_n$, $\B_n$ and ${\bm 0}_n$ denote the points 
$(a_1,\ldots,a_n)\in \mathbb{R}^n$, $(b_1,\ldots,b_n)\in \mathbb{R}^n$, and $(0,0,\ldots,0)\in \mathbb{R}^n$, respectively.
\begin{defn}
  For $\A_n,\B_n\in \mathbb{R}^n$, the Euclidean distance of $\A_n$ and $\B_n$ is defined by$$\rho(\A_n,\B_n):=\sqrt{\sum_{i=1}^n (a_i-b_i)^2}.$$
\end{defn}
\begin{defn}
For $\A_n\in \mathbb{R}^n$, let $B_{\A_n}(r)$ be the open ball which centered in $\A_n$ with radius $r$, that is $$B_{\A_n}(r):= \{\B_n\in \mathbb{R}^n\mid \rho (\A_n,\B_n) < r\}.$$
\end{defn}
\begin{defn}
Let $f\in \R[\xx_n]$, the set of real zeros of $f$ is denoted by $\zero(f)$.
Let $L$ be a subset of $\R[\xx_n]$. Define
$$\zero(L)=\{\A_n\in\R^n|\forall f\in L, f(\A_n)=0\}.$$
The elements of $\zero(L)$ are the common real zeros of $L$. If $L=\{f_1,\ldots,f_m\}$, $\zero(L)$ is also denoted by $\zero(f_1,\ldots,f_m)$.
\end{defn}
\begin{defn}
The {\em level} of $f\in \mathbb{R}[\xx_n]$ is the largest $j$ such that $\deg(f,{x_j})>0$ where $\deg(f,x_j)$ is the degree of $f$ with respect to $x_j$.
  The {\em level} of $f\in \mathbb{R}[\xx_n,k]$ is the largest $j$ such that $\deg(f,{x_j})>0$ and the level is zero if all $x_i$s do not appear in $f$.
  \end{defn}
\begin{defn}
    For a polynomial set $L\subseteq \mathbb{R}[\xx_n]$ or $L\subseteq \mathbb{R}[\xx_n,k]$, $L^{i}$ is the set of polynomials in $L$ of level $i$.
 \end{defn}

\begin{defn}
    For $\A_n,\B_n\in \mathbb{R}^n$, we denote by $\A_n\B_n$ the segment $\A_n\rightarrow \B_n$. For $m$ points $\A_n^1$, $\ldots$, $\A_n^m$, we denote by $\A_n^1 \rightarrow \A_n^2\rightarrow \cdots \rightarrow \A_n^m$ the broken line through $\A_n^1,\ldots,\A_n^m$ in turn.
\end{defn}

The following two lemmas are well-known results.
\begin{lem}\label{lem:res} Let $ f,g\in \R[\xx_n]$, there exist nonzero $p, q \in \R[\xx_n]$ such that $pf+qg = \Res(f,g,x_n)$ with $\deg(p,x_n)<\deg(g,x_n)$ and $\deg(q,x_n)<\deg(f,x_n)$, where $\Res(f,g,x_n)$ is the resultant of $f$ and $g$ with respect to $x_n$.
\end{lem}
\begin{pf}
See, for example \citep{Cox2005}.
\end{pf}

\begin{lem}\label{lem:zhou}Let $f(\xx_n)\in\R[\xx_n]$ and $r$ be a real positive number. If $f(\A_n)=0$ for all $\A_n\in B_{{\bm 0}_n}(r)$, then $f(\xx_n)\equiv 0$.
\end{lem}
\begin{pf}
See, for example \citep{Marshall2008}.
\end{pf}


\begin{lem}\label{lem:trans} For $f,g \in \R[\xx_n]$, if $f$ and $g$ are coprime in $\R[\xx_n]$, then after any linear invertible transform, $f$ and $g$ are still coprime in $\R[\xx_n]$, namely for $A \in GL_n(\mathbb{R}), B_n\in \R^n$, $\xx^{*T}_n = A\xx^T_n + B^T_n$, then $\gcd(f(\xx^{*}_n),g(\xx^{*}_n)) = 1$ in $\R[\xx_n]$.
\end{lem}
\begin{pf}If $\gcd(f(\xx^{*}_n),g(\xx^{*}_n)) = h(\xx_n)$ and $h$ is not a constant. Then $h(A^{-1}(\xx_n-B^T_n))$ is a non-trivial common divisor of $f$ and $g$ in $\mathbb{R}[\xx_n]$, 
which is a contradiction.
\end{pf}

\begin{lem}\label{lem:disc}Suppose $f,g \in \R[\xx_n]$ and $\gcd(f,g) = 1$ in $\R[\xx_n]$. For any $\A_{n-1}\in \mathbb{R}^{n-1}$ and $r>0$,
there exists $\A_{n-1}'\in \mathbb{R}^{n-1}$ such that $\rho(\A_{n-1},\A_{n-1}')<r$ and for all $a_n' \in \mathbb{R}$, $(\A_{n-1}',a_n') \notin \zero(f,g)$.
\end{lem}
\begin{pf}Otherwise, there exist $\A_{n-1}^0=(a_1^0,\ldots,a_{n-1}^0) \in \mathbb{R}^{n-1}$ and $r_0>0$, such that for any $\A_{n-1}^1=(a_1^1,\ldots,a_{n-1}^1)$ satisfying $\rho(\A_{n-1}^0,\A_{n-1}^1)<r_0$, there exists an $a_n^1 \in \mathbb{R}$ such that $f(\A_{n-1}^1,a_n^1) = g(\A_{n-1}^1,a_n^1) = 0$.
Thus $\Res(f,g,x_n)= 0$ at every point of $B_{\A_{n-1}^0}(r_0)$. From Lemma \ref{lem:zhou}, we get that $\Res(f,g,x_n)\equiv 0$, 
meaning $\gcd(f,g)$ is non-trivial, which is impossible.
\end{pf}

\begin{defn}
Let $f(x) =c_lx^l+\cdots+c_0\in \R[x]$ with $c_l \neq 0.$
The {\em discriminant} of $f(x)$ is
$$\discrim(f,x)=c_l^{2l-2}\prod_{i<j}{(z_i-z_j)}^2,$$
where $z_i$ $(i=1,\ldots,l)$ are the complex roots of the equation $f(x)=0$.\par
The following well-known equation shows the relationship between $\discrim(f,x)$ and $\Res(f,\frac{\partial}{\partial x}f,x)$,
$$c_l\discrim(f,x)=(-1)^{\frac{l(l-1)}{2}}\Res(f,\frac{\partial}{\partial x}f,x).$$
Suppose the coefficients of $f$ are given parametrically as polynomials in $\xx_n$. If the {\em leading coefficient} $\lc(f,x)=c_l\not\equiv 0$, the discriminant of $f(\xx_n,x)$ can be written as
$$\discrim(f,x)= (-1)^{\frac{l(l-1)}{2}}\left(
    \begin{array}{cccccccc}
    1       & c_{l-1}    &c_{l-2}     &  \cdots    & c_j    &\cdots     \\
    0       & c_l        &  c_{l-1}   &  \cdots    & c_{j+1}&\cdots   \\
    0       & 0          & c_l        &  \cdots    &c_{j+2} & \cdots   \\
    \vdots  & \vdots     &  \vdots    &  \ddots    & \vdots & \ddots   \\
    l       &(l-1)c_{l-1}&(l-2)c_{l-2}&  \cdots   &  jc_j   & \cdots   \\
    0       &lc_l        &(l-1)c_{l-1}& \cdots     & (j+1)c_{j+1}&\cdots \\
    0       & 0          &lc_l        & \cdots    & (j+2)c_{j+2}    & \cdots\\
    \vdots  & \vdots     & \vdots     & \ddots     &\vdots     &  \ddots
    \end{array}
    \right).$$
    If $c_l=c_{l-1}=0$ at point $\A_n$, from the above expression, $\discrim(f,x)=0$ at this point.
\end{defn}

\begin{lem}\label{lem:weiss}\citep{weiss}
Let $f(x)\in\R[x]$ be a monic squarefree polynomial of degree $l$, the sign of its discriminant is $(-1)^{\frac{l-r}{2}}$, where $r$ is the number of its real roots.
\end{lem}
It is clear that the conclusion of the above lemma still holds when $\lc(f,x)$ is positive.
\begin{lem}\label{lem:discrim}
Given a polynomial $f(\xx_n,x_{n+1})\in \R[\xx_n,x_{n+1}]$, say
$$f(\xx_n,x_{n+1})=\sum_{i=0}^lc_ix_{n+1}^{i}, c_l \not\equiv 0,$$
where $c_i (i=0,\ldots,l)$ is a polynomial in $\xx_n$. Let $U$ be an open set in $\mathbb{R}^{n}$.
If $f(\xx_n,x_{n+1})\ge0$ on $U\times \R$, then $l$ is even and
$$(-1)^{\frac{l}{2}}\discrim(f,x_{n+1})\ge0 \quad\text{and}\quad \lc(f,x_{n+1})\ge0 \quad\text{for all} \quad \A_n\in U.$$
\end{lem}
\begin{pf}
Since $f$ is positive semi-definite for any given $\A_n\in U$, $\lc(f,x_{n+1})$ is positive semi-definite on $U$
and $l$ is even. If $c_l > 0$ at $\A_n$ and $f(\A_n, x_{n+1})$ is squarefree, then $$(-1)^{\frac{l}{2}}\discrim(f(\xx_n,x_{n+1}),x_{n+1})\mid _{\xx_n=\A_n}=(-1)^{\frac{l}{2}}\discrim(f(\A_n,x_{n+1}),x_{n+1}) > 0$$ by Lemma \ref{lem:weiss}. Otherwise, either $c_l = 0$ at $\A_n$ which suggests $c_{l-1} = 0$ at $\A_n$, or $c_l > 0$ at $\A_n$ and $f(\A_n,x_{n+1})$ is not squarefree. In both cases we can deduce $$(-1)^{\frac{l}{2}}\discrim(f(\xx_n,x_{n+1}),x_{n+1}) = 0$$ at $\A_n$.
That completes the proof.
\end{pf}
Before we go further, we would like to give a remark on the coefficient ring of polynomials.

\begin{rem}\label{re:R}
Although most of the theorems of this paper are valid for $\R[\xx_n]$, we restrict ourselves to $\Z[\xx_n]$ when we design algorithms because they need effective factorization and real root isolation. Actually, suppose $\rr$ is a subring of $\R$ and takes $\Z$ as a subring. If $\rr[\xx_n]$ admits effective factorization and $\rr[x]$ admits effective real root isolation, all the algorithms in this paper are effective. Two examples of such rings are ${\mathbb Q}$ and the field of real algebraic numbers. In the following, we use $\rr$ to denote such a ring.
\end{rem}

\begin{defn}\label{de:sqrfree}
Suppose $h$$\in \rr[\xx_n]$ can be factorized in $\rr[\xx_n]$ as:
  $$h=a{h_1}^{i_1}{h_2}^{i_2}\ldots {h_m}^{i_m},$$
where $a\in\rr$, $h_i (i=1,\ldots,m)$ are pairwise different irreducible monic polynomials (under a suitable ordering) with degree greater than or equal to one in $\rr[\xx_n]$. Define
  $$\sqrfree(h)=\prod_{i=1}^m{h_i}.$$
  If $h$ is a constant, let $\sqrfree(h)=1.$
\end{defn}

\begin{lem}\label{lem:xia}
Given a real polynomial $f$ with real parameters, say
$$f(\mathbf{c},x)=c_mx^m+c_{m-1}x^{m-1}+\ldots+c_0,$$
where $\mathbf{c}=(c_m,\ldots,c_0)$ is real parameter.
Let $R(\mathbf{c})=\sqrfree(\Res(f,f',x))$. If $\mathbf{s_1}$ and $\mathbf{s_2}$ are two points in the same connected component of parameter space $R(\mathbf{c})\neq0$, then
$f(\mathbf{s_1},x)$ and $f(\mathbf{s_2},x)$ have the same number of real roots $y_1(\mathbf{s_1})<y_2(\mathbf{s_1})<\ldots<y_d(\mathbf{s_1})$ and $y_1(\mathbf{s_2})<y_2(\mathbf{s_2})<\ldots<y_d(\mathbf{s_2})$. Moreover, $y_i(\mathbf{c})$$(i=1,\ldots,d)$ is continuous in the connected component.
\end{lem}
\begin{pf}
See for example \citet{xia}.
\end{pf}

In the following, we introduce some basic concepts and results of CAD. The reader is referred to \citet{collins1}, \citet{Hong},  \citet{McCallum1,McCallum2}, \citet{brown} and \citet{xiaorong} for a detailed discussion on the properties of CAD and Open CAD.

\begin{defn}\citep{collins1,McCallum1}
  An $n$-variate polynomial $f(\xx_{n-1},x_{n})$ over the reals is said to be {\em delineable} on a subset $S$ (usually connected) of $\mathbb{R}^{n-1}$ if
(1) the portion of the real variety of $f$ that lies in the cylinder $S\times \mathbb{R}$ over $S$ consists of the union of the graphs of some $t\ge0$ continuous functions $\theta_1<\cdots<\theta_t$ from $S$ to $\mathbb{R}$; and
(2) there exist integers $m_1,\ldots,m_t\ge1$ such that for every $a\in S$, the multiplicity of the root $\theta_i(a)$ of $f(a,x_n)$ (considered as a polynomial in $x_n$ alone) is $m_i$.
\end{defn}

\begin{defn}\citep{collins1,McCallum1}
 In the above definition, the $\theta_i$ are called the real root functions of $f$ on $S$, the graphs of the $\theta_i$ are called the $f$-sections over $S$, and the regions between successive $f$-sections are called $f$-sectors.
\end{defn}

\begin{thm}\label{thm:McCallum}\citep{McCallum1,McCallum2}
Let $f(\xx_n,x_{n+1})$ be a polynomial in $\mathbb{R}[\xx_n,x_{n+1}]$ of positive degree and $\discrim(f,x_{n+1})$ is a nonzero polynomial. Let $S$ be a connected submanifold of $\mathbb{R}^n$ on which $f$ is degree-invariant and does not vanish identically, and in which $\discrim(f,x_{n+1})$ is order-invariant. Then $f$ is analytic delineable on $S$ and is order-invariant in each $f$-section over $S$.
\end{thm}
Based on this theorem, McCallum proposed the projection operator \MCproj, which consists of the discriminant of $f$ and all coefficients of $f$.

\begin{thm}\citep{brown}\label{thm:Brown}
 Let $f(\xx_n,x_{n+1})$ be an $(n+1)$-variate polynomial of positive degree $m$ in the variable $x_{n+1}$ with $\discrim(f,x_{n+1})\neq0$. Let $S$ be a connected submanifold of $\mathbb{R}^n$ where $\discrim(f,x_{n+1})$ is order-invariant, the leading coefficient of $f$ is sign-invariant, and such that $f$ vanishes identically at no point in $S$. $f$ is degree-invariant on $S$.
\end{thm}
Based on this theorem, Brown obtained a reduced McCallum projection in which only leading coefficients and discriminants appear. 
\begin{defn}\citep{brown}
 Given a polynomial $f\in \rr[\xx_{n}]$ of level $n$,
  the Brown projection operator for $f$ is
  $$\Bproj(f,x_{n})=\Res(\sqrfree(f),\frac{\partial (\sqrfree(f))}{\partial x_{n}}, x_{n}).$$
If $L$ is a polynomial set and the level of any polynomial in $L$ is $n$, then
  \begin{align*}
  \Bproj(L,x_{n})=&\cup_{f\in L}\{\Res(\sqrfree(f),\frac{\partial (\sqrfree(f))}{\partial x_{n}}, x_{n})\}\bigcup\\
  &\cup_{f,g\in L, f\neq g}\{\Res(\sqrfree(f),\sqrfree(g),x_{n})\}.
  \end{align*}
\end{defn}

\begin{algorithm}[ht]
\caption{{\tt Bprojection} \citep{brown}}\label{Bproj}
\begin{algorithmic}[1]
\Require{A polynomial $f(\xx_n)\in \Z[\xx_n]$.}
\Ensure{A projection factor set $F$.}
\State $F:=\{f(\xx_n)\}$;
\For{$i$ from $n$ downto 2}
\State $F:=F\bigcup \{\Bproj(F^{i},x_{i})\}$; ($F^{i}$ is the set of polynomials in $F$ of level $i$).
\EndFor
\State\textbf{return} $F$
\end{algorithmic}
\end{algorithm}

Open CAD is a modified CAD construction algorithm, which was named in
Rong Xiao's Ph.D. thesis \citep{xiaorong}. In fact, Open CAD is similar to the Generic Cylindrical Algebraic Decomposition (GCAD) proposed by \citet{Strzebonski} and was used in DISCOVERER \citep{xia2000} for real root classification. For convenience, we describe the framework of the Open CAD here.

For a polynomial $f(\xx_n)\in \rr[\xx_n]$, an Open CAD defined by $f(\xx_n)$ is a set of rational sample points in $\R^n$ obtained through the following three phases:
(1) Projection. Use the Brown projection operator (Algorithm \ref{Bproj}) on $f(\xx_n)$;
(2) Base. Choose one rational point in each of the open intervals defined by the real roots of $F^1$ (see Algorithm \ref{Bproj});
(3) Lifting. Substitute each sample point of $\R^{i-1}$ for $\xx_{i-1}$ in $F^i$ and then, by the same method as Base phase, choose rational sample points for $F^i(x_i)$.\par


\section{The Successive Resultant Method}\label{se:4}
The Successive Resultant Method (SRes) was introduced in \citep{yang2001} without a proof. The method can be used for solving Problem 2 of this paper, i.e., problem of global optimization.

For a polynomial $f(\xx_n)$ in Problem 2, the SRes method first applies Algorithm \ref{Bproj} on polynomial $f(\xx_n)-K$ to get a polynomial $g(K)$. Suppose $g(K)$ has $m$ distinct real roots $k_i (1\le i\le m)$. Then computes $m+1$ rational numbers $p_i (0\le i\le m)$ such that $k_i\in (p_{i-1}, p_i)$. Finally, substitutes each $p_i$ in turn for $K$ in $f(\xx_n)-K$ to check if $f(\xx_n)-p_i\ge 0$ holds for all $\xx_n$. If $p_j$ is the first such that $f(\xx_n)-p_j\ge 0$ does not hold, then $k_j$ is the infimum (let $k_0=-\infty, k_{m+1}=+\infty$). To check if $f(\xx_n)-p_i\ge 0$ holds for all $\xx_n$, the SRes method applies Brown's projection on $f(\xx_n)-p_i$ and choose sample points by Open CAD in the lifting phase.

The SRes method is formally described as Algorithm \ref{SRES} and we prove its correctness in the rest part of this section.


\begin{algorithm}[ht]
\caption{\SRes\ (Successive Resultant Method, \citet{yang2001,xia})}
 \label{SRES}
\begin{algorithmic}[1]
\Require{A squarefree polynomial $f\in \Z[\xx_n]$.}
\Ensure{The supremum of $k\in\mathbb{R}$, such that $\forall \A_n\in \mathbb{R}^n$, $f(\A_n)\ge k$. If there doesn't exist such $k$, then returns $-\infty$.}
\State $g:=f-k$ ($g$ is viewed as a polynomial in $k\prec x_1\prec\cdots\prec x_n$);
\State $F:={\tt Bprojection}(g)$ ($F^{i}$ is the set of polynomials in $F$ of level $i$. Here $F^{i}$ has no more than one polynomial, we denote this polynomial by $F_i$.);
\State $C_0:=$an Open CAD of $\mathbb{R}$ defined by $F_{0}(k)$ (Suppose $C_0=\bigcup_{i=0}^m \{p_i\}$, $p_i \in(k_i,k_{i+1})$, where $k_i$ $(1\le i\le m)$ are the real roots of $F_0$ and $k_0=-\infty$, $k_{m+1}=+\infty$.);
\For {$l$ from $0$ to $m$}
\For {$i$ from $1$ to $n$}
\State {$C_{li}$:= an Open CAD of $\mathbb{R}^i$ defined by $\bigcup_{j=1}^i F_j(\xx_j,p_l)$};
\EndFor
\If {there exists a sample point $\A_n$ in $C_{ln}$ such that $f(\A_n)-p_l<0$}
\State\textbf{return} $k_{l}$
\EndIf
\EndFor
\State \textbf{return} $-\infty$
\end{algorithmic}
\end{algorithm}

 \begin{rem}
   If $g(\xx_n)\ge0$ for all $\xx_n\in \R^n$, Algorithm \ref{SRES} can also be applied to compute $\inf\{\frac{f(\xx_n)}{g(\xx_n)}|\xx_n\in\R^n\}$. We just need to replace $F:={\tt Bprojection}(f-k)$ of Line 2 by
$F:={\tt Bprojection}(f-kg)$. The proof of the correctness is the same.
 \end{rem}

The following lemma can be inferred from the results of \citep{McCallum2} and \citep{brown}, {\it i.e.}, $f$ is delineable over the maximal connected regions defined by $\Bproj(f,x_{n})\neq0$. We give a new proof here.

\begin{lem}\label{lem:zhuyinli}\citep{McCallum2,brown}
Let $F_i, F_{i-1}$ be as in Algorithm \ref{SRES}.
Let $U$ be a connected component of $F_{i-1}\neq0$ in $\R^{i}$ and $y_i^1(\gamma) < y_i^2(\gamma)<\cdots< y_i^m(\gamma)$ be all real roots of $F_{i}(\gamma,x_i) = 0$ for any given $\gamma\in U.$
Then for all $\alpha, \beta\in U$, $\alpha\times(y_i^{j-1}(\alpha), y_i^j(\alpha))$ and $\beta\times(y_i^{j-1}(\beta), y_i^j(\beta)) (j = 2,3,\ldots,m)$
are in the same connected component of $F_{i}\neq0$ in $\R^{i+1}$. 
\end{lem}
\begin{pf}
         For $\alpha \in U$, let $\varepsilon = \min\limits_{2\le i \le m}\mid y_i(\alpha) - y_{i-1}(\alpha)\mid$, by Lemma \ref{lem:xia}, $\exists \delta >0, $ such that $\forall \alpha' \in B(\alpha, \delta)$, $\max\limits_{1\le i \le m}\mid y_i(\alpha) - y_{i}(\alpha')\mid < \frac{\varepsilon}{6}$.\par
         Consider segment $(\alpha, \frac{y_{j-1}(\alpha) + y_{j}(\alpha)}{2})\rightarrow(\alpha', \frac{y_{j-1}(\alpha') + y_{j}(\alpha')}{2})$ where $\alpha' \in B(\alpha, \delta)$. For any point $(\alpha'', y)$ on the segment, we have
    \begin{displaymath}
    \begin{aligned}
                 &\mid y-y_s(\alpha'')\mid\\
                =&\mid (y_s(\alpha'')-y_s(\alpha))+(y_s(\alpha)-\frac{y_{j-1}(\alpha) + y_j(\alpha)}{2})
                   +(\frac{y_{j-1}(\alpha)+y_j(\alpha)}{2}-y)\mid\\
        \ge&\mid y_s(\alpha)-\frac{y_{j-1}(\alpha) + y_j(\alpha)}{2}\mid
                    -\mid y_s(\alpha'')-y_s(\alpha)\mid-\mid\frac{y_{j-1}(\alpha)+y_j(\alpha)}{2}-y\mid\\
        \ge&\frac{\varepsilon}{2}-\frac{\varepsilon}{6}
                    -\mid\frac{y_{j-1}(\alpha)+y_j(\alpha)}{2}-\frac{y_{j-1}(\alpha')+y_j(\alpha')}{2}\mid\\
        \ge&\frac{\varepsilon}{2}-\frac{\varepsilon}{6}-\frac{\varepsilon}{6}\\
                >&0
    \end{aligned}
    \end{displaymath}
         So the points satisfying $F_{i} = 0$ are not on the segment.\par
         Therefore, for any points $ r_1 \in (y_{j-1}(\alpha), y_j(\alpha)), r_2 \in (y_{j-1}(\alpha'), y_j(\alpha')), (\alpha' \in B_{\alpha}(\delta))$,
         the points satisfying $F_{i} = 0$ are not on the broken line $(\alpha, r_1)\rightarrow(\alpha, \frac{y_{j-1}(\alpha) + y_j(\alpha)}{2})\rightarrow(\alpha', \frac{y_{j-1}(\alpha') + y_j(\alpha')}{2})\rightarrow(\alpha', r_2)$.\par
         Hence we know that for any $\alpha \in U$, there exists $\delta >0$ such that for any point $\alpha' \in B_{\alpha}(\delta)$ and $2\le s\le m$, $\alpha \times(y_{s-1}(\alpha), y_s(\alpha))$ and $\alpha' \times(y_{s-1}(\alpha'), y_s(\alpha'))$ are in the same connected component of $F_{i}\neq0$ in $\R^{i+1}$.
         For all $\alpha, \beta\in U$, there exists a path $\gamma:[0,1]\rightarrow U$ that connects $\alpha$ and $ \beta$. Due to the compactness of the path,
         there are finitely many open sets $B_{\alpha_t}(\delta_t)$ covering $\gamma([0,1])$ with $\alpha_t \in \gamma([0,1]),
          \forall \alpha' \in B_{\alpha_t}(\delta_t), 2\le j\le m, \alpha \times(y_{j-1}(\alpha), y_j(\alpha))$ and
          $\alpha' \times(y_{j-1}(\alpha'), y_j(\alpha'))$ are in the same connected component of $F_{i}\neq0$. Since the union of these open sets are connected, the lemma is proved.
\end{pf}
\begin{rem}\label{rem:opencadpro}
By the above Lemma, in Algorithm \ref{SRES}, for any two points $p_l$, $p_l'$ $\in(k_l,k_{l+1})$, their corresponding sample points obtained through the Open CAD lifting phase are in the same connected component of $F_n\neq0$ in $\mathbb{R}^{n+1}$.
Since at least one sample point can be taken from every highest dimensional cell via the Open CAD lifting phase,
the set of the corresponding sample points of $p_l$ obtained through the Open CAD lifting phase, $C_{ln}$ in Algorithm \ref{SRES}, contains at least one point from every connected component $U$ of $F_n(\xx_n,k)\neq0$, in which $U\bigcap (\R^n\times (k_{l-1},k_{l}))\neq\emptyset$.
\end{rem}

\begin{thm}
   The Successive Resultant Method is correct.
\end{thm}
 \begin{pf}
Let notations be as in Algorithm \ref{SRES}.
 If there exists a $k' \in (k_i, k_{i+1})$, such that $ F_n(\xx_n, k') \ge 0$ for all $\xx_n\in \mathbb{R}^n$, then by Lemma \ref{lem:zhuyinli}, for any $k \in (k_i, k_{i+1})$, $F_n(\xx_n, k) \ge 0$ for all $\xx_n \in \mathbb{R}^n$ (since their corresponding sample points obtained through the Open CAD lifting phase are in the same connected component of $F_n(\xx_n,k)\neq0$ in $\mathbb{R}^{n+1}$). Therefore,
 for any $k \in [k_i, k_{i+1}]$, $F_n(\xx_n, k) \ge 0$ for all $\xx_n \in \mathbb{R}^n$. The global optimum $k$ will be found by checking whether $\forall \A_n \in \mathbb{R}^n, F_n(\A_n,p_i) \ge 0$ holds where $p_i$ is the sample point of $(k_i,k_{i+1})$. Since Algorithm \ref{SRES} ensures that at least one point is chosen from every connected component of $F_{n}(\xx_n,p_i)\neq0$ in $\R^{n}$, the theorem is proved.
\end{pf}

\section{Solving Problem 1 via simplified CAD projection}\label{se:5}

To improve the efficiency of CAD based methods for solving Problem 1, {\it i.e.}, proving or disproving $f(\xx_n)\ge 0$, we propose a new projection operator called \Nproj. The operator has been illustrated by a simple example in Section \ref{se:2}. In this section, we give a formal description of our method for solving Problem 1 based on \Nproj\ and prove its correctness.

\subsection{Notations}

\begin{defn}\label{de:sqrfree1}
Suppose $h\in \rr[\xx_n]$ can be factorized in $\rr[\xx_n]$ as:
  $$h=al_{1}^{2j_1-1}\ldots l_t^{2j_t-1}{h_1}^{2i_1}\ldots {h_m}^{2i_m},$$
where $a\in \rr$, $h_i$$(i=1,\ldots,m)$ and $l_j$$(i=1,\ldots,t)$ are pairwise different irreducible monic polynomials 
(under a suitable ordering) with degree greater than or equal to one in $\rr[\xx_n]$. Define
  \begin{align*}
  &\sqrfree_1(h)=\{l_i,i=1,2,\ldots,t\},\\
  &\sqrfree_2(h)=\{h_i,i=1,2,\ldots,m\}.
  \end{align*}
  If $h$ is a constant, let $\sqrfree_1(h)=\{1\},$ $\sqrfree_2(h)=\{1\}.$
\end{defn}

  \begin{defn}
  Suppose $f\in \rr[\xx_{n}]$ is a polynomial of level $n$. Define   
  \begin{align*}
  & {\rm Oc}(f,x_n)=\sqrfree_1(\lc(f,x_{n})),\ {\rm Od}(f,x_n)=\sqrfree_1(\discrim(f,x_{n})),\\
  & {\rm Ec}(f,x_n)=\sqrfree_2(\lc(f,x_{n})),\ {\rm Ed}(f,x_n)=\sqrfree_2(\discrim(f,x_{n})),\\
  & {\rm Ocd}(f,x_n)={\rm Oc}(f,x_n)\cup {\rm Od}(f,x_n),\ {\rm Ecd}(f,x_n)={\rm Ec}(f,x_n)\cup {\rm Ed}(f,x_n).
  \end{align*}
  The {\em secondary} and {\em principal parts} of the new projection are defined as
  $$\Nproj_1(f,x_{n})={\rm Ocd}(f,x_n),$$
   $$\Nproj_{2}(f,x_{n})=\{\prod_{g\in {\rm Ecd}(f,x_{n})\setminus {\rm Ocd}(f,x_{n})}{g}\}.$$
  If $L$ is a set of polynomials of level $n$, define
  $$\Nproj_1(L,x_{n})=\cup_{g\in L}{\rm Ocd}(g,x_{n}),$$
  $$\Nproj_{2}(L,x_{n})=\bigcup_{g\in L}\{\prod_{h\in {\rm Ecd}(g,x_n)\setminus \Nproj_1(L,x_{n})}{h}\}.$$
\end{defn}

\subsection{Algorithm}


By Theorem \ref{thm:4} (see Section 5.3 for details), 
the task of proving $f(\xx_n)\ge 0$ on $\R^n$ can be accomplished by (1) computing sample points of $\Nproj_{2}(f,x_n)\neq0$ in $\R^{n-1}$ and checking $f(\alpha,x_n)\ge0$ on $\R$ for all sample points $\alpha$; and (2) proving all the polynomials in $\Nproj_{1}(f,x_n)$ are positive semi-definite on $\mathbb{R}^{n-1}$. For (1), typical CAD based methods, {\it e.g.}, Open CAD, can be applied. For (2), we can call this procedure recursively. Now the idea of our algorithm \Proineq\ is clear and is formally described here.

\begin{algorithm}[hb]
\caption{\Proineq}\label{pro}
\begin{algorithmic}[1]
\Require{A polynomial $f(\xx_n) \in \Z[\xx_n]$\ (monic under a suitable ordering)}
\Ensure{Whether or not $f(\xx_n) \ge0$ on $\R^n$}
\If {$f$ is a constant}
\If {$f\ge0$ }
\Return  \textbf{true}
\Else {
\Return  \textbf{false}}
\EndIf
\Else
\If {$f$ is reducible in $\Z[\xx_n]$}
\For {$g$ in $\sqrfree_1(f)$}
\If {$\Proineq(g)=$\textbf{false}}
\Return  \textbf{false}
\EndIf
\EndFor
\EndIf
\State $L_1:=\Nproj_1(f,x_n)$
\State $L_2:=\Bprojection(\Nproj_2(f,x_n))\bigcup \{f(\xx_n)\}$
\For {$g$ in $L_1$}
\If {$\Proineq(g)=$\textbf{false}}
\Return  \textbf{false}
\EndIf
\EndFor
\For {$i$ from 1 to $n$}
\State $C_i:=$ An Open CAD of $\mathbb{R}^i$ defined by $\cup_{j=1}^i L_2^{j}$ {(If $i=n-1$, we require that for any sample point $\A_{n-1}$ in $C_{n-1}$, $\A_{n-1}\notin\bigcup_{h\in L_1}\zero(h)$)}
\EndFor
\If{there exists an $\A_{n}\in C_{n}$ such that $f(\A_{n})<0$}
\Return \textbf{false}
\EndIf
\State\textbf{return} \textbf{true}
\EndIf
\end{algorithmic}
\end{algorithm}

To give the readers a picture of how our new projection operator is different from existing CAD projection operators, we give
Algorithm \ref{nproj} here, which returns all possible polynomials that may appear in the projection phase of Algorithm \ref{pro}.
\begin{algorithm}[ht]
\caption{$\Nproj$}
\label{nproj}
\begin{algorithmic}[1]
\Require{A polynomial $f(\xx_n)\in \Z[\xx_n]$.}
\Ensure{Two projection factor sets containing all possible polynomials that may appear in the projection phase of Algorithm \ref{pro}.}
\State $L_1:=\sqrfree_1(f)$;
\State $L_2:=\{\}$;
\For{$i$ from $n$ downto 2}
\State $L_2:=L_2\bigcup \Nproj_{2}(L_1^{i},x_i)\bigcup \cup_{g\in L_2^{i}}\Bproj(g,x_i)$; (Recall that $L^{i}$ is the set of polynomials in $L$ of level $i$.)
\State $L_1:=L_1\bigcup \Nproj_1(L_1^{i},x_i)$;
\EndFor
\State\textbf{return} $(L_1,L_2).$
\end{algorithmic}
\end{algorithm}


\begin{rem}\label{re:nproj}
For polynomial $P(x_1,\ldots,x_{n-1},x_n)=p(x_1,\ldots,x_{n-1},x_n^2)$ $(\deg(P,x_n)\ge2,n\ge2),$ the resultant of $P$ and $P_{x_n}'$ with respect to $x_n$ is (may differ from a constant)
$$p(x_1,\ldots,x_{n-1},0)\Res(p,p_{x_n}',x_{n})^2.$$
If $p(x_1,\ldots,x_{n-1},0)$ is not a square, the set $\Nproj_1(P,x_n)$ is not empty and thus the scale of $\Nproj(P)$ is smaller than that of $\Bprojection(P)$.

If for any polynomial $f\in\Z[x_1,\ldots,x_n]$, the iterated discriminants of $f$ always have odd factors and are reducible (for generic $f$ or for most polynomials, it is quite likely), then for $n\ge3$, the the scale of $\Nproj(f)$ is always strictly smaller than that of $\Bprojection(f)$.
\end{rem}

\subsection{The correctness of Algorithm \Proineq}
\begin{thm}\label{thm:1}Let $f(\xx_n)$ and $g(\xx_n)$ be coprime in $\R[\xx_n]$. For any connected open set $U$ in $\mathbb{R}^n$, the open set $V = U\backslash \zero(f,g)$ is also connected.
\end{thm}
This theorem plays an important role in our proof. It can be proved by the fact that closed and bounded semi-algebraic set is semi-algebraically triangulable \citep{Bochnak1998} and Alexander duality. Here we give an elementary proof.
\begin{pf}
For any two points $\alpha$, $\beta$ in $V$, we only need to prove that there exists a path
$\gamma(t):[0,1]\rightarrow V$ such that $\gamma(0) = \alpha, \gamma(1) = \beta.$
Choose a path $\gamma_U$ that connects $\alpha$ and $\beta$ in $U$. Notice that $U$ is an open set, so for any $X_n \in
\gamma_U$, there exists $\delta_{X_n} > 0$ such that $U\supset B_{X_n}(\delta_{X_n})$. Since $\gamma_U$ is compact and
$\bigcup B_{X_n}(\delta_{X_n})$ is an open covering of $\gamma_U$, there exists an $m \in \mathbb{N}$, such that
$\bigcup \limits_{k = 1}\limits^{m} B_{X_n^k}(\delta_{X_n^k}) \supset \gamma_U$ and $\alpha \in
B_{X_n^1}(\delta_{X_n^1})$, $\beta \in B_{X_n^m}(\delta_{X_n^m})$, $B_{X_n^i}(\delta_{X_n^i})\bigcap B_{X_n^{i+1}}(\delta_{X_n^{i+1}})\neq \emptyset$ $(i=1,2,\ldots,m-1)$. Now we only need to prove that for every $k$,
$B_{X_n^k}(\delta_{X_n^k})\backslash \zero(f,g)$ is connected. If this is the case, we can find $k$ paths $\gamma_1$, $\gamma_2$, $\ldots,$ $\gamma_k$ with $\gamma_1(0)=\alpha$, $\gamma_1(1)\in B_{X_n^1}(\delta_{X_n^1})\bigcap B_{X_n^{2}}(\delta_{X_n^{2}})$, $\gamma_{i+1}(0)=\gamma_{i}(1)$, $\gamma_{i+1}(1)\in B_{X_n^i}(\delta_{X_n^i})\bigcap B_{X_n^{i+1}}(\delta_{X_n^{i+1}})$ $(i=1,2,\ldots,m-1)$, $\gamma_m(1)=\beta$. Let $\gamma$ be the path: $[0,1]\rightarrow U$ which satisfies $\gamma([\frac{j-1}{m},\frac{j}{m}])=\gamma_j([0,1])$ $(j=1,2,\ldots,m)$, then $\gamma$ is the path as desired.

Choose $a$, $b\in B_{X_n^k}(\delta_{X_n^k})\backslash \zero(f,g)$. There exists an affine coordinate transformation $T$ such that $T(B_{X_n^k}(\delta_{X_n^k}))=B_{\mathbf{0}_n}(1)$ and $\overrightarrow{T(a)T(b)}$ and $\overrightarrow{(\mathbf{0}_{n-1},1)}$ are parallel. Thus the first $n-1$ coordinates of $T(a)$ and $T(b)$ are the same.
Let $T(a)=(Y_{n-1},a')$, $T(b)=(Y_{n-1},b')$. Without loss of generality, we assume that $a'>b'$.

In the new
coordinate, $f$ and $g$ become $T(f)$ and $T(g)$, respectively. $B_{\mathbf{0}_n}(1)$ is an open set and
$T(a),T(b)\notin \zero(T(f),T(g))$, so there exists $r>0$ such that the cylinder
$B_{Y_{n-1}}(r)\times[b',a'] \subseteq B_{\mathbf{0}_n}(1)$, $B_{T(a)}(r)\bigcap\zero(T(f),T(g))=\emptyset$ and
$B_{T(b)}(r)\bigcap$ $\rm{Zero}$ $(T(f),T(g))=\emptyset$.
By Lemma \ref{lem:trans}, $T(f)$ and $T(g)$ are coprime in $\mathbb{R}[\xx_n]$. So by Lemma \ref{lem:disc}, there
exists $X_{n-1}'\in B_{Y_{n-1}}(r)$, such that for any $x_n \in \mathbb{R}$, $ (X_{n-1}',x_n) \notin \rm{Zero}$
$(T(f),T(g))$. Thus the broken line
$T(a)\rightarrow(X_{n-1}',a')\rightarrow(X_{n-1}',b')\rightarrow T(b)$ is a path that connects $T(a)$ and $T(b)$
in $B_{\mathbf{0}_n}(1)\backslash \zero(T(f),T(g))$.
The theorem is proved.
\end{pf}

\begin{prop}\label{tuilun:1} Suppose $U\subseteq \R^n$ is a connected open set, $f,g \in \R[\xx_n]$, $\gcd(f,g) = 1$ in $\R[\xx_n]$ and for all $X_n \in U$, $f(X_n)g(X_n) \ge 0$. Then either $f(X_n)\ge0,g(X_n)\ge0$ for all $X_n \in U$ or $f(X_n)\le0,g(X_n)\le0$ for all $X_n \in U$. Similarly, if for all $X_n \in U$, $f(X_n)g(X_n) \le 0$, then either $f(X_n)\ge0,g(X_n)\le0$ for all $X_n \in U$ or $f(X_n)\le0,g(X_n)\ge0$ for all $X_n \in U$.
\end{prop}
\begin{pf}If not, there exist $ X_{n}^1$, $X_{n}^2 \in U$, such that $f(X_{n}^1)\le0$, $g(X_{n}^1)\le0$ and $f(X_{n}^2)\ge0$, $g(X_{n}^2)\ge0$. By Theorem \ref{thm:1}, $U\backslash\zero(f,g)$ is connected. So we can choose a path $\gamma$ that connects $X_n^1$ with $X_n^2$ and $\gamma\bigcap \zero(f,g)=\emptyset$.
Consider the sign of $f+g$ on $\gamma$. Since the sign is different at $X_n^1$ and $X_n^2$, by Mean Value Theorem we know there exists $X_n^3$ on $\gamma$ such that $f(X_n^3)+g(X_n^3)=0$. From the condition we know that $f(\xx_n)g(\xx_n)\ge0$, hence $X_n^3\in \zero(f,g)$, which contradicts the choice of $\gamma$.

The second part of the proposition can be proved similarly.
\end{pf}

The following proposition is an easy corollary of Proposition \ref{tuilun:1}.
\begin{prop}\label{tuilun:alg3}
 Let $f\in \rr[\xx_n]$ be a monic (under a suitable ordering) polynomial of level $n$, the necessary and sufficient condition for $f(\xx_n)$ to be positive semi-definite on $\mathbb{R}^n$ is, for any polynomial $g\in\sqrfree_1(f)$, $g$ is positive semi-definite on $\mathbb{R}^n$.
\end{prop}
\begin{prop}\label{tuilun:2}
  Suppose $f\in \R[\xx_n]$ is a non-zero squarefree polynomial and $U$ is a connected open set of $\mathbb{R}^{n}$. If
  $f(\xx_n)$ is semi-definite on $U$, then $U\backslash \zero(f)$ is also a connected open set.
\end{prop}
\begin{pf}
Without loss of generality, we assume $f(\xx_{n})\ge0$ on $U$. Since $f$ is non-zero, we only need to consider the case that the level of $f$ is non-zero. Let $i>0$ be the level of $f$ and consider $f$ as a polynomial of $x_i$. Because $f(\xx_n)\ge 0$ on $U$, we know ${\rm Zero}(f)\bigcap U = {\rm Zero}(f,f'_{x_i})\bigcap U.$ Otherwise, we may assume there exists a point $X_n^0=(x_1^0,\ldots,x_n^0)\in U$ such that $f(X_n^0)=0$ and $f'_{x_i}(X_n^0)>0$. Thus, there exists $r$ such that $\forall X_n\in B_{X_n^0}(r)$, $f'_{x_i}(X_n)>0$.
Let
$F(x_i)=f(x_1^0,\ldots,x_{i-1}^0$, $x_i$, $x_{i+1}^0,\ldots,x_{n}^0)$. The Taylor series of $F$ at point $x_i^0$ is
$$F(x_i)=F(x_i^0)+(x_i-x_i^0)F'_{x_i}(x_i^0+\theta(x_i-x_i^0)),$$
where $\theta\in(0,1)$. Let $x_i^0>x_i^1>x_i^0-r$, then $F(x_i^1)<0$, which contradicts the definition of $F$.

If $f$ is irreducible in $\R[\xx_n]$, $f$ and $f'_{x_i}$ are coprime in $\R[\xx_n]$. Thus $U\backslash \zero(f,f'_{x_i})$ is connected by Theorem \ref{thm:1}. So $U\backslash \zero(f)$ is a connected open set.

If $f$ is reducible in $\R[\xx_n]$, let $f=a\prod_{t=1}^j f_t$, in which $a\in\R$ and all $f_t (t=1,\ldots,j)$ are irreducible monic polynomials (under a suitable ordering) in $\R[\xx_n]$ , then $U\backslash \zero(f) =U\backslash \bigcup_{t=1}^j\zero(f_t)$ is a connected open set.
The proposition is proved.
\end{pf}
\begin{thm}\label{thm:2} Given a positive integer $n\ge2$. Let $f\in \rr[\xx_n]$ be a non-zero squarefree polynomial and $U$ be a connected component of $\Nproj_{2}(f,x_{n})\neq0$ in $\R^{n-1}$. If the polynomials in $\Nproj_{1}(f,x_n)$ are semi-definite on $U$, then $f$ is delineable on $V=U\backslash \bigcup_{h\in \Nproj_{1}(f,x_n)}\zero(h)$.
\end{thm}
\begin{pf}
 According to Theorem \ref{thm:McCallum} and Theorem \ref{thm:Brown}, $f$ is delineable over the connected component of $\Res(f,f_{x_n}',x_n)\neq0$. By Proposition \ref{tuilun:2}, $V=U\backslash \bigcup_{h\in \Nproj_{1}(f,x_n)}\zero(h)$ is a connected open set. Thus, $f$ is delineable on $V$.
 \end{pf}


\begin{thm}\label{thm:3}
Given a positive integer $n\ge2$. Let $f\in\rr[X_{n}]$ be a squarefree polynomial of level $n$ and $U$ a connected open set of $\Nproj_{2}(f,x_{n})\neq0$ in $\R^{n-1}$. The necessary and sufficient condition for $f(\xx_n)$ to be semi-definite on $U\times \R$ is the following two conditions hold.\\
$(1)$The polynomials in $\Nproj_{1}(f,x_n)$ are semi-definite on $U$;\\
$(2)$There exists a point $\alpha\in U\backslash \bigcup_{h\in \Nproj_{1}(f,x_n)}\zero(h)$, $f(\alpha,x_n)$ is semi-definite on $\mathbb{R}$.
\end{thm}
\begin{pf}
$\Longrightarrow:$ By Lemma \ref{lem:discrim}, $\discrim(f,x_n)$ is semi-definite on $U$. Thus by Proposition \ref{tuilun:1},
the polynomials in $\Nproj_{1}(f,x_n)$ are semi-definite on $U$. It is obvious that $f(\alpha,x_n)$ is semi-definite on $\mathbb{R}$.

$\Longleftarrow$: If the polynomials in $\Nproj_{1}(f,x_n)$ are semi-definite on $U$, by Theorem \ref{thm:2}, $f$  is delineable on the connected open set $V=U\backslash \bigcup_{h\in \Nproj_{1}(f,x_n)}\zero(h)$.
From that $f(\alpha,x_n)$ is semi-definite on $\mathbb{R}$, we know that $f(\xx_n)$ is semi-definite on $U\times \R$.
\end{pf}

The following theorem is an easy corollary of the above theorem.
\begin{thm}\label{thm:4}
Given a positive integer $n\ge2$. Let $f\in \rr[\xx_n]$ be a squarefree monic (under a suitable ordering) polynomial of level $n$, the necessary and sufficient condition for $f(\xx_n)$ to be positive semi-definite on $\mathbb{R}^n$ is the following two conditions hold. \\
$(1)$ The polynomials in $\Nproj_{1}(f,x_n)$ are positive semi-definite on $\mathbb{R}^{n-1}$;\\
$(2)$ For every connected components $U$ of $\Nproj_{2}(f,x_n)\neq0$, there exists a point $\alpha\in U$, and $\alpha$ is not a zero of any polynomial in $\Nproj_{1}(f,x_n)$, such that $f(\alpha,x_n)\ge0$ on $\R$.
 \end{thm}

\begin{thm}
 Algorithm \ref{pro} is correct.
\end{thm}

\begin{pf}
  By Proposition \ref{tuilun:alg3}, we only need to consider the case that $f(\xx_n)$ is irreducible in $\Z[\xx_n]$. When $n=1$, it is obvious that Algorithm \ref{pro} is correct. \par
  We prove the theorem by induction on the level of $f$. Now, suppose that Algorithm \ref{pro} is correct for every polynomial $h$ of level less than or equal to $n-1$. If $f$ is positive semi-definite on $\R^n$, by Theorem \ref{thm:4}, the polynomials in $\Nproj_{1}(f,x_n)$ are positive semi-definite on $\mathbb{R}^{n-1}$. By induction, $\Proineq$ returns true for all these polynomials. Since $f$ is positive semi-definite, $f(X_n^1)\ge0$ for all sample points obtained in $\Proineq(f)$. Thus $\Proineq(f)$ returns true. If $f$ is not positive semi-definite, by Theorem \ref{thm:4}, there are two possible cases. \par
  (1) There exists at least one polynomial in $\Nproj_{1}(f,x_n)$ which is not positive semi-definite on $\mathbb{R}^{n-1}$. Since the level of this polynomial is less than $n$, for this case, by induction, Algorithm \ref{pro} returns false.\par
  (2) There exists a connected open set $U$ of $\Nproj_{2}(f,x_n)\neq0$, a point $\alpha\in U$ where $\alpha$ is not a zero of any polynomial in $\Nproj_{1}(f,x_n)$ and a point $a\in \R$ such that $f(\alpha,a)<0$. By (1), we can assume the polynomials in $\Nproj_{1}(f,x_n)$ are positive semi-definite on $\mathbb{R}^{n-1}$. So,  by Theorem \ref{thm:2}, $f$ is delineable on $V=U\backslash \bigcup_{h\in \Nproj_{1}(f,x_n)}\zero(h)$. Thus, for any $\beta\in U$, there exists a point $b\in \R$ such that $f(\beta,b)<0$. By the lifting property of Open CAD, in Algorithm \ref{pro}, there exists a sample point $X_{n-1}^0\in C_{n-1}$ with $X_{n-1}^0\in V$. Thus there exists $c\in \R$ such that $(X_{n-1}^0,c)\in C_{n}$, $f(X_{n-1}^0,c)<0$. Algorithm \ref{pro} returns false in this case.
\end{pf}

\section{Solving Problems 2 and 3}\label{se:6}



\subsection{Problem 3}

Recall that Problem 3 is: For $f\in \R[\xx_n,k]$, find all $k_0\in \mathbb{R}$ such that $f(\xx_n,k_0)\ge 0$ on $\R^n$. Since this is a typical QE problem, any CAD based methods can be applied. Under a suitable ordering on variables, {\it e.g.}, $k\prec x_1\prec\cdots\prec x_n$,
by CAD projection, one can obtain a polynomial in $k$, say $g(k)$. Assume $k_1<\cdots<k_m$ are the real roots of $g(k)$ and $p_j\in (k_{j-1}, k_j) (1\le j\le m+1)$ are rational sample points in the $m+1$ intervals where $k_0=-\infty, k_{m+1}=+\infty$. Then checking whether or not $f(\xx_n,k_i)\ge 0 (1\le i\le m)$ and $f(\xx_n,p_j)\ge 0 (1\le j\le m+1)$ on $\R^n$ will give the answer. Namely, if there exist $p_j$ such that $f(\xx_n,p_j)\ge 0$ then $(k_{j-1}, k_j)$ should be output. If $f(\xx_n,k_i)\ge 0$ for some $k_i$, $\{k_i\}$ should be output.

Thus, a natural idea for improving efficiency is to apply the new projection operator \Nproj\ instead of Brown's projection in the above procedure.
In this subsection, we first show by an example why \Nproj\ cannot be applied directly to Problem 3. Then we propose an algorithm based on \Nproj\ for solving Problem 3 and prove its correctness.

\begin{ex}
Find all $k\in\R$ such that $$(\forall x,y\in\R) f(x,y,k)=x^2+y^2-k^2\ge0.$$
If we apply \Nproj\ directly (with an ordering $k\prec x\prec y$), we will get $$\Nproj(f)=(\{f(x,y,k),x-k,x+k,1\}, \{1\}).$$
Because $L_2=\{1\}$, there is only one sample point with respect to $k$, say $k_0=0$. Substituting $k_0$ for $k$ in $f(x,y,k)$, we check whether
$(\forall x,y\in\R) x^2+y^2\ge0.$
This is obviously true. So, it leads to a wrong result:
$(\forall k,x,y\in\R) x^2+y^2-k^2\ge0.$
\end{ex}

The reason for the error is that $(x-k)(x+k)$ will be a square if $k=0$. The point $k=0$ can be found by computing the resultant $\Res(x-k,x+k,x)$ which is avoided by \Nproj\ since $x-k\in L_1$ and $x+k\in L_1$.

This example indicates that, if we use \Nproj\ to solve Problem 3, we have to consider some ``bad" values of $k$ at which some odd factors of $\sqrfree_1(f)$ may become some new even factors. In the following, we first show that such ``bad" values of $k$ are finite and propose an algorithm for computing all possible ``bad" values. Then we give an algorithm for solving Problem 3, which handles the ``bad" values and the ``good" values of $k$ obtained by \Nproj\ separately.


\begin{defn}\label{de:bad}
Let $f(\xx_n,k)\in \Z[\xx_n,k]$ and $(L_1,L_2)=\Nproj(f(\xx_n,k))$ with the ordering $k\prec x_1\prec\cdots\prec x_n$. If $\alpha\in \R$ satisfying that
\begin{enumerate}
\item there exist two different polynomials $g_1,g_2\in \sqrfree_1(f(\xx_n,k))$ such that $g_1|_{k=\alpha}$ and $g_2|_{k=\alpha}$ have non-trivial common factors in $\R[\xx_n]$; or
\item there exist an $i (2\le i \le n)$, a polynomial $g\in L_1^i$ and two different polynomials $g_1,g_2\in \Nproj_1(g, x_i)$ such that $g_1|_{k=\alpha}$ and $g_2|_{k=\alpha}$ have non-trivial common factors in $\R[\xx_n]$; or
\item there exists a polynomial $g\in L_1$ such that $g|_{k=\alpha}$ has non-trivial square factors in $\R[\xx_n]$,
\end{enumerate}
then $\alpha$ is called a {\em bad value} of $k$. The set of all the bad values is denoted by ${\rm Bad}(f,k)$.
\end{defn}

For two coprime multivariate polynomials with parametric coefficients, the problem of finding all parameter values such that the two polynomials have non-trivial common factors at those parameter values is very interesting. We believe that there should have existed some work on this problem. However, we do not find such work in the literature. So, we use an algorithm in \citep{qian}. The detail of improvements on the algorithm is omitted.

\begin{algorithm}[ht]
\caption{{\tt BK}}
\label{alg:qian1}
\begin{algorithmic}[1]
\Require{Two coprime polynomials $f(\xx_n, k), g(\xx_n, k)\in \Z[\xx_n,k]$ and $k$.}
\Ensure{$B$, a finite set of polynomials in $k$.}
\State $B:=\emptyset$;
\State $r:=\Res(f,g,k)$; Let $S$ be the set of all irreducible factors of $r$.
\State Let $X={\rm indets}(S);$\ \  ($X$ is the set of variables appearing in $S$)
\While{$X\ne \emptyset$ }
\State Choose a variable $x\in X$ such that the cardinal number of\\ $T=\{p\in S|~ x ~\mbox{appears in}~ p\}$ is the biggest;
\State $h:=\Res(f,g,x)$;
\State $B:= B \cup \{q(k)|~ q(k) ~\mbox{is irreducible} ~\mbox{and divides}~ h \}$;
\State $S:=S\setminus T;~ X:={\rm indets}(S)$;
\EndWhile
\State\textbf{return} $B$
\end{algorithmic}
\end{algorithm}

\begin{algorithm}[ht]
\caption{$\NKproj$}
\label{alg:nkproj}
\begin{algorithmic}[1]
\Require{A polynomial $f(\xx_n,k)\in \Z[\xx_n]$ and an ordering $k\prec x_1\prec\cdots\prec x_n$.}
\Ensure{Two projection factor sets as in Algorithm \ref{nproj} and a set of polynomials in $k$.}
\State $L_1:=\sqrfree_1(f)$;\ $L_2:=\{\}$;\ $B:=\emptyset$;
\For{$i$ from $n$ downto 1}
\State $L_2:=L_2\bigcup \Nproj_{2}(L_1^{i},x_i)\bigcup \cup_{g\in L_2^{i}}\Bproj(g,x_i)$;
\For {$h\in L_1^i$}
\State $L_{1h}:=\Nproj_1(h,x_i)$;
\State $B:=B\cup {\tt BK}(h,\frac{\partial}{\partial x_i}h,k)$;
\State $B:=B\bigcup\cup_{h_1\ne h_2\in L_{1h}}{\tt BK}(h_1,h_2,k)$;
\State $L_1:=L_1\cup L_{1h}$;
\EndFor
\EndFor
\State\textbf{return} $(L_1,L_2,B).$
\end{algorithmic}
\end{algorithm}

It is not hard to prove the following lemmas.
\begin{lem}\label{lem:qian}\citep{qian}
${\tt BK}(f,g,k)\supseteq \{\alpha\in\R|~ \gcd(f(\xx_n, \alpha), g(\xx_n, \alpha)) ~\mbox{is non-trivial}\}.$
\end{lem}

\begin{lem}\label{lem:kmax}
Let notations be as in Algorithm \ref{alg:nkproj}.
\begin{enumerate}
\item The first two outputs, $L_1$ and $L_2$, are the same as $\Nproj(f(\xx_n,k))$ with the ordering $k\prec x_1\prec\cdots\prec x_n$.
\item $\bigcup_{h\in B}\zero(h)\supseteq {\rm Bad}(f)$. Thus, ${\rm Bad}(f,k)$ is finite.
\item If $k_0$ is not a bad value and $f(\xx_n,k_0)\ge0$ on $\R^n$, then for any $h\in \sqrfree_1(f)$, $h(\xx_n,k_0)$ is semi-definite on $\R^n$.
\end{enumerate}
\end{lem}
\begin{pf}
(1) and (2) are obvious. For (3), because $k_0\notin {\rm Bad}(f)$, $g_1(\xx_n,k_0)$ and $g_2(\xx_n,k_0)$ are coprime in $\Z[\xx_n]$ for any $g_1\ne g_2\in\sqrfree_1(f)$. Since $f(\xx_n,k_0)\ge0$, by Proposition \ref{tuilun:1}, for any $h\in \sqrfree_1(f(\xx_n,k))$, $h(\xx_n,k_0)$ is semi-definite on $\R^n$.
\end{pf}

\begin{algorithm}[ht]
\caption{{\tt Findk}}
\label{findk}
\begin{algorithmic}[1]
\Require{A polynomial $f(\xx_n,k)\in \mathbb{Z}[\xx_n,k]$.}
\Ensure{A set $FK_f$.}
\State $FK_f:=\{\}$;
\State $(L_1,L_2,B):=\NKproj(f)$; \ (with an ordering $k\prec x_1\prec\cdots\prec x_n$)
\State Suppose $\zero(L_1^{0}\cup L_2^{0})=\cup_{i=1}^{m}\{k_i\}$ with $k_1<\cdots<k_m.$ Let $k_0=-\infty$, $k_{m+1}=+\infty$. 
\For {$l$ from $1$ to $m+1$}
\State{Choose a sample point $p_l \in(k_{l-1},k_{l})\setminus \bigcup_{h\in B}\zero(h)$;}
\State $v:=1$;
\For {$i$ from $1$ to $n$}
\State $C_{li}$:=an Open CAD of $\mathbb{R}^i$ defined by $\cup_{j=1}^i L_1^{j}\mid_{k=p_l}\bigcup \cup_{j=1}^i L_2^{j}\mid_{k=p_l}$;
\If {$i=n$ and there exists $X_n\in C_{ln}$ such that $f(X_n,p_l)<0$}
\State $v:=0$;
\Else \If{there exist $X_i^1,X_i^2\in C_{li}$ and $g\in L_1^i$ such that $g(X_i^1,p_l)g(X_i^2,p_l)<0$}
\State $v:=0$;
\State{\textbf{break}}
\EndIf
\EndIf
\EndFor
\If {$v=1$}
\State $FK_f:=FK_f \bigcup(k_{l-1},k_{l})$;
\EndIf
\EndFor
\For {$\alpha$ in $\{k_1,\ldots,k_m\}\cup\bigcup_{h\in B}\zero(h)\setminus FK_f$}
\If {$\Proineq(f(\xx_n,\alpha))=$\textbf{true}}
\State $FK_f:=FK_f \bigcup\{\alpha\}$;
\EndIf
\EndFor
\State\textbf{return} $FK_f$
\end{algorithmic}
\end{algorithm}

\begin{lem}\label{lem:findk}
Let $f(\xx_n,k)\in \mathbb{Z}[\xx_n,k]$ and $(L_1,L_2,B)=\NKproj(f)$. Suppose $\zero(L_1^{0}\cup L_2^{0})=\cup_{i=1}^{m}\{k_i\}$ with $k_1<\cdots<k_m,$  $k_0=-\infty$, $k_{m+1}=+\infty$ and, for every $l (1\le l\le m+1)$, $p_l \in(k_{l-1},k_{l})\setminus \bigcup_{h\in B}\zero(h)$. Denote by $C_{li}$ an Open CAD of $\mathbb{R}^i$ defined by $\cup_{j=1}^i L_1^{j}\mid_{k=p_l}\bigcup \cup_{j=1}^i L_2^{j}\mid_{k=p_l}$. If there exists an $l$($1\le l \le m+1$), such that \\
$(1)$ $\forall X_n\in C_{ln}$, $f(X_n,p_l)\ge0$; and\\
$(2)$ $\forall i (0\le i\le n-1)\forall g\in L_1^i\forall X_i^1,X_i^2\in C_{li}$, $g(X_i^1, p_l)g(X_i^2, p_l)\ge 0$,\\
then
for any $0\le i\le n$ and $g_i(\xx_i,k)$ in $L_1^i$, $g_i(\xx_i,k)$ is semi-definite on $\R^i\times (k_{l-1},k_{l})$.
\end{lem}
\begin{pf}
    We prove it by induction on $i$. When $i=0$, the conclusion is obvious. When $i=1$, by Theorem \ref{thm:3}, it is also true.
    Assume the conclusion is true when $i=j-1 (j\ge2)$. For any polynomial $g_j(\xx_j,k)$ in $L_1^j$,
    notice that $\Nproj_1(g_j)\subseteq L_1^{j-1}$, $\Nproj_2(g_j)\subseteq L_2^{j-1}$.
    By the assumption of induction, we know that every polynomial in $\Nproj_1(g_j)$ is semi-definite on $\R^{j-1}\times (k_{l-1},k_{l}).$ By Theorem \ref{thm:3}, $g_j(\xx_j,k)$ is semi-definite on $\R^j\times (k_{l-1},k_{l})$. That finishes the induction.
\end{pf}

\begin{thm}\label{thm:findk}
  The output of Algorithm \ref{findk}, $FK_f$, is $\{\alpha\in\R|\forall X_n\in \mathbb{R}^n,f(X_n,\alpha)\ge0\}$.
\end{thm}
\begin{pf}
Denote $\{\alpha\in\R|\forall X_n\in \mathbb{R}^n,f(X_n,\alpha)\ge0\}$ by $K_f$. \par
We first prove that $FK_f\subseteq K_f$. Suppose $(k_{l-1},k_{l})\subseteq FK_f$. Since $\sqrfree_1(f)\in L_1^n$, the semi-definitenss of $f$ on $\R^{n}\times (k_{l-1},k_{l})$ follows from Lemma {\rm \ref{lem:findk}}. Because we check the positive definiteness of $f$ on sample points, $(k_{l-1},k_{l})\subseteq K_f$.

We then prove that $K_f\subseteq FK_f$. It is sufficient to prove that if there exists $k'\in(k_{l-1},k_{l})\setminus \bigcup_{h\in B}\zero(h)$ such that $\forall X_n\in \mathbb{R}^n,f(X_n,k')\ge0$, then $(k_{l-1},k_{l})\in FK_f$. 

It is obviously true when $n=1$. When $n\ge2$, for any $g_n\in \sqrfree_1(f)$, since $f(\xx_n,k')$ is semi-definite and $k'\notin \bigcup_{h\in B}\zero(h)$, $g_n(\xx_n,k')$ is semi-definite by Lemma \ref{lem:kmax}. For any
\[g_{n-1}(\xx_{n-1},k)\in \Nproj_{1}(g_n(\xx_n,k),x_n)={\rm Oc}(g_n,x_n)\cup{\rm Od}(g_n,x_n),\]
we have \[\sqrfree_1(g_{n-1}(\xx_{n-1},k'))\subseteq \Nproj_{1}(g_n(\xx_n,k'),x_n)\] because $k'\notin\bigcup_{h\in B}\zero(h).$ By Theorem \ref{thm:4}, $g_{n-1}(\xx_{n-1},k')$ is semi-definite on $\R^{n-1}$. Hence, for any polynomial $g_{n-1}(\xx_{n-1},k)$ in $L_1^{n-1}$, $g_{n-1}(\xx_{n-1},k')$ is semi-definite. In a similar way, we know that for any $1\le j\le n-1$ and any polynomial $g_{j}(\xx_j,k)$ in $L_1^{j}$, $g_{j}(\xx_j,k')$ is semi-definite on $\mathbb{R}^{j}$. 
Therefore, for any $0\le i\le n$ and any polynomial $g_i(\xx_i,k)$ in $L_1^i$, $g_i(\xx_i,k)$ is semi-definite on $\R^i\times (k_{l-1},k_{l})$  by Lemma \ref{lem:findk}. Hence, no matter what point $p_l\in (k_{l-1},k_{l})$ is chosen as the sample point of this open interval, $(k_{l-1},k_{l})$ will be in the output of Algorithm \ref{findk}, {\it i.e.}, $(k_{l-1},k_{l})\in FK_f$. The proof is completed.
\end{pf}

\subsection{Problem 2}
For solving the global optimum problem (Problem 2), 
we only need to modify the algorithm \Findk\ a little and get the algorithm \Findinf.

\begin{algorithm}[ht]
\caption{\Findinf}
\label{findkmax}
\begin{algorithmic}[1]
\Require{A squarefree polynomial $f\in \Z[\xx_n]$.}
\Ensure{$k\in \mathbb{R}$ such that $k=\inf_{\xx_n\in\mathbb{R}^n}f(\xx_n)$. }
\State $(LI_1,LI_2):=\Nproj(f(\xx_n)-k)$\ \ (with an ordering $k\prec x_1\prec\cdots\prec x_n$)
\State Suppose $\zero(LI_1^{0}\cup LI_2^{0})=\cup_{i=1}^{m}\{k_i\}$ with $k_1<\cdots<k_m.$ Let $k_0=-\infty$, $k_{m+1}=+\infty$.
\For {$l$ from $1$ to $m+1$}
\State{Choose a sample point $p_l$ of $(k_{l-1},k_{l})$}
\State $v:=1$;
\For {$i$ from $1$ to $n$}
\State $C_{li}$:=an Open CAD of $\mathbb{R}^i$ defined by $\cup_{j=1}^i LI_1^{j}\mid_{k=p_l}\bigcup \cup_{j=1}^i LI_2^{j}\mid_{k=p_l}$;
\If {$i=n$ and there exists $X_n\in C_{ln}$ such that $f(X_n)-p_l<0$}
\State $v:=0$;
\Else
\If{there exist $X_i^1,X_i^2\in C_{li},g\in LI_1^i$ such that $g(X_i^1,p_l)g(X_i^2,p_l)<0$}
\State $v:=0$;
\State{\textbf{break}}
\EndIf
\EndIf
\EndFor
\If {$v=0$}
\State\textbf{return} $k_{l-1}$
\EndIf
\EndFor
\end{algorithmic}
\end{algorithm}

\begin{thm}\label{th:findmax}
  The output of Algorithm \ref{findkmax} is the global infimum $\inf f(\mathbb{R}^n)$.
\end{thm}
\begin{pf}
We only need to prove that if there exists $k'\in(k_{l-1},k_{l})$ such that $f(\xx_n)\ge k'$ on $\R^n$, then $f(\xx_n)\ge k_l$ on $\R^n$.

The result is obviously true when $n=1$. When $n\ge2$, we can find a ``good" value $k''\in (k_{l-1},k')\setminus {\rm Bad}(f-k,k)$ because the bad values are finite according to Lemma \ref{lem:kmax}. Since $f(\xx_n)\ge \bar{k}$ for $\bar{k}\in(k_{l-1},k')$, $f(\xx_n)-k''\ge 0$. Then, by Lemma \ref{lem:kmax} (3), $h(\xx_n,k'')$ is semi-definite on $\R^n$ for any $h\in \sqrfree_1(f(\xx_n)-k)$. In a similar way, we know that for any $1\le j\le n-1$ and any polynomial $g_{j}(\xx_j,k)$ in $LI_1^{j}$, $g_{j}(\xx_j,k'')$ is semi-definite on $\mathbb{R}^{j}$. Therefore, for any $0\le i\le n$ and any polynomial $g_i(\xx_i,k)$ in $LI_1^i$, $g_i(\xx_i,k)$ is semi-definite on $\R^i\times (k_{l-1},k_{l})$ by Lemma \ref{lem:findk}. Hence, $f(\xx_n)-k$ is positive semi-definite on $\R^n\times(k_{l-1},k_{l})$ by Theorem \ref{thm:3}.
\end{pf}
\begin{rem}
For $f,g\in \R[\xx_n]$, if $g(\xx_n)\ge0$ on $\R^n$, Algorithm \Findinf\ can also be applied to compute $\inf\{\frac{f(\xx_n)}{g(\xx_n)}|\xx_n\in\R^n\}$. We just need to replace $(LI_1,LI_2):=\Nproj(f(\xx_n)-k)$ of Line 1 by
$(LI_1,LI_2):=\Nproj(f(\xx_n)-kg(\xx_n))$.
\end{rem}

\section{Examples}\label{se:7}
We haven't made any complexity analysis on our new algorithms. We believe that the complexity is still doubly exponential but we do not know how to prove it yet.
In this section, we report the performance of Algorithms \Findinf, \Findk\ and \Proineq\ on several non-trivial examples.
Since our main contribution is an improvement on the CAD projection for solving those three special problems, we only make some comparison with other CAD based tools on these examples.
Algorithm \Findinf\ will be compared with the algorithm \SRes. The program \Proineq\ we implemented using Maple will be compared with the function PartialCylindricalAlgebraicDecomposition (\PCAD) of RegularChains package in Maple15, function FindInstance in Mathematica9, and QEPCAD B.

Because we do not have Mathematica and QEPCAD B installed in our computer, we ask others' help. So the computations were performed on different computers. FindInstance (\FI) was performed on a laptop with Inter Core(TM) i5-3317U 1.70GHz CPU and 4GB RAM. QEPCAD B (\QEPCAD) was performed on a laptop with Intel(R) Core(TM) i5 3.20GHz CPU and 4GB RAM.
 The other computations were performed on a laptop with Inter Core2 2.10GHz CPU and 2GB RAM.


We show the different results of projection of Algorithm \Findinf\ and Algorithm \SRes\ by Example \ref{ex:1}.
\begin{ex}\label{ex:1}
Compute $\inf_{x,y,z\in\mathbb{R}}G(x,y,z)$, where
$$G=\frac{(x^2-x+1)(y^2-y+1)(z^2-z+1)}{(xyz)^2-xyz+1}.$$
Let $f=(x^2-x+1)(y^2-y+1)(z^2-z+1), g=(xyz)^2-xyz+1.$ Since $g\ge0$ for any $x,y,z\in \R$,
this problem can be solved either by Algorithm \Findinf\ or by Algorithm \SRes.

If we apply Algorithm \Findinf, after
$\Nproj(f-kg)$ with an ordering $k\prec z\prec y\prec x$, we will get a polynomial in $k$,
$$P=(k-\frac{2}{3})(k^2+6k-3)(k-\frac{279}{256})k(k-1)(k-\frac{3}{4})(k-\frac{9}{16})(k-9),$$
which has $9$ distinct real roots. After sampling and checking signs, we finally know that the maximum $k$ is the real root of $k^2+6k-3$ in $(\frac{1}{4}, \frac{1}{2})$.

If we apply Algorithm \SRes, after $\tt Bprojection$$(f-kg)$ with an ordering $k\prec z\prec y\prec x$, we  will get a polynomial in $k$,
\begin{align*}
F=&\frac{1}{614656}(614656k^4-4409856k^3+11013408k^2-11477376k+4021893)\cdot\\
                 &(k^4-294k^3+1425k^2-2277k+1089)(k-\frac{9}{4})\cdot P,
\end{align*}
which has 14 distinct real roots. After sampling and checking signs, we finally know that the maximum $k$ is the real root of $k^2+6k-3$ in $(\frac{1}{4}, \frac{1}{2})$.

Obviously, the scale of projection with the new projection is smaller. The polynomial in $k$ calculated through the successive resultant method has three extraneous factors.
\end{ex}

\begin{ex}\citep{han} Prove
$$F(\xx_n,n)=\prod_{i=1}^n(x_i^2+n-1)- n^{n-2}(\sum_{i=1}^n x_i)^2\ge0\ \mbox{on}\ \R^n.$$
When $n=3,4,5,6,7$, we compared \Proineq, \FI, \PCAD, \QEPCAD\ in the following table. Hereafter $>$3000 means either the running time is over 3000 seconds or the software is failure to get an answer.
\begin{center}
\begin{tabular}{lllll}
\hline$n\diagup$Time(s)& \Proineq \hspace{1cm}&  \FI\hspace{1cm}& \PCAD\hspace{1cm}&\QEPCAD\\
\hline
$3$        &0.063&0.015&0.078&0.020\\
$4$        &0.422&0.062&0.250&0.024\\
$5$        &0.875&2.312&2.282&0.372\\
$6$        &4.188&$>$3000&$>$3000&$>$3000\\
$7$        &$>$3000&$>$3000&$>$3000&$>$3000\\
\hline
\end{tabular}
\end{center}
When $n=3,4,5,6$, we compared the number of polynomials in the projection sets of \Bproj\ with \Nproj (under the same ordering) as well as the number of sample points need to be chosen through the lifting phase under these two projection operators.
\begin{center}
\begin{tabular}{llllll}
\hline$n$ \hspace{1cm}        & \Bprojection \hspace{1.4cm}               & &      & \Nproj \hspace{1.4cm}               &                    \\
\cline{2-3} \cline{5-6}
   &\# polys& \# points&&\# polys&\# points\\
\hline
$3$        &11&4 & &8&3 \\
$4$        &22& 10 & &12&3\\
$5$        &88& 36 & &18&5 \\
$6$        &Unknown& Unknown& &32&15\\
\hline
\end{tabular}
\end{center}
\end{ex}
\begin{ex}
Decide the nonnegativity of $G(n,k)$
$$G(n,k)=(\sum_{i=1}^n x_i^2)^2-k\sum_{i=1}^n x_{i}^3x_{i+1},$$
where $x_{n+1}=x_1$.

In the following table, (T) means that the corresponding program outputs $G(n,k)\ge 0$ on $\R^n$. (F) means the converse.
\begin{center}
\begin{tabular}{lllll}
\hline$(n,k)\diagup$Time(s)& \Proineq \hspace{1cm}&  \FI\hspace{1cm}& \PCAD\hspace{1cm}&\QEPCAD\\
\hline
$(3,3)$        &0.047(T)&0.031(T)&0.078(T)&0.032(T)\\
$(4,3)$        &0.171(T)&284.484(T)&0.891(T)&196.996(T)\\
$(5,3)$        &244.188(T)&$>$3000&$>$3000&$>$3000\\
$(6,3)$        &$>$3000&$>$3000&$>$3000&$>$3000\\
$(4,k_1)$      &13.782(F)&5638.656(F)& 24.656(F)&$>$3000\\
\hline
\end{tabular}
\end{center}
where $k_1=\frac{227912108939855024517609}{75557863725914323419136}$.
By Algorithm \Findk, for the case $n=4$, we can find the maximum value of $k$ satisfying the following inequality
$$(\forall (x_1,x_2,x_3,x_4)\in\R^4)\quad G(4,k)=(\sum_{i=1}^4 x_i^2)^2-k\sum_{i=1}^4 x_{i}^3x_{i+1}\ge0$$
is the real root in $(3,\frac{7}{2})$ 
of $$800000k^8-29520000 k^6+311367675k^4-422100992k^2-5183373312=0.$$

\begin{center}
\begin{tabular}{llllll}
\hline$n$ \hspace{1cm}        & \Bprojection \hspace{1.4cm}               & &      & \Proineq \hspace{1.4cm}               &                    \\
\cline{2-3} \cline{5-6}
&\# polys&\# points&&\# polys&\# points\\
\hline
$3$        &5&10 & &4&5 \\
$4$        &6& 4 & &5&2\\
$5$        &Unknown& Unknown & &16&20 \\
\hline
\end{tabular}
\end{center}
\end{ex}

The following example was once studied by \citet{parrilo}.
\begin{ex}
$$\forall X_{3m+2}\in\R^{3m+2},B(x)=(\sum_{i=1}^{3m+2}x_i^2)^2-2\sum_{i=1}^{3m+2}x_i^2\sum_{j=1}^mx_{i+3j+1}^2\ge0,$$
\begin{center}
\begin{tabular}{lllll}
\hline$3m+2\diagup$Time(s)& \Proineq \hspace{1cm}&  \FI\hspace{1cm}& \PCAD\hspace{1cm}&\QEPCAD\\
\hline
$5$        &0.297&0.109&0.265&0.104\\
$8$        &27.218&$>$3000&$>$3000&$>$3000\\
$11$       &$>$3000&$>$3000&$>$3000&$>$3000\\
\hline
\end{tabular}
\end{center}
\begin{center}
\begin{tabular}{llllll}
\hline$3m+2$ \hspace{0.5cm}        & \Bprojection \hspace{1.4cm}               & &      & \Proineq \hspace{1.4cm}               &                    \\
\cline{2-3} \cline{5-6}
&\# polys&\# points&&\# polys&\# points\\
\hline
$5$        &13&96 & &10&88 \\
$8$        &Unknown& Unknown & &27&6720\\
\hline
\end{tabular}
\end{center}
\end{ex}

The above examples demonstrate that in terms of proving inequalities, among CAD based methods, Algorithm \Proineq\ is efficient and our new algorithms can work out some examples which could not be solved by other existing general CAD tools.

Further improvements on the projection and lifting phase are our ongoing work.

\begin{ack}
The authors would like to convey their gratitude to Professor McCallum and all the three referees who provided their valuable comments, advice and suggestion, which help improve this paper greatly on not only the presentation but also the technical details.

The authors also would like to thank all members in the symbolic computation seminar of Peking University, including Liyun Dai who simplified the proof of Proposition \ref{tuilun:1}, Xiaoxuan Tang, Zhenghong Chen, Ting Gan and Xiaoyu Qian, who together checked the proofs.
\end{ack}

\end{document}